\documentclass[preprint]{aastex}

\shorttitle{\lq\lq DARK\rq\rq GRB 080325 IN A DUSTY MASSIVE GALAXY AT $z \sim 2$}
\shortauthors{Hashimoto et al.}
\begin{document}
\title{\lq\lq DARK\rq\rq GRB 080325 IN A DUSTY MASSIVE GALAXY AT $z \sim 2$\altaffilmark{*}}
\author{
T. Hashimoto\altaffilmark{1}, K. Ohta\altaffilmark{1}, 
K. Aoki\altaffilmark{2}, I. Tanaka\altaffilmark{2}, 
K. Yabe\altaffilmark{1}, N. Kawai\altaffilmark{3}, 
W. Aoki\altaffilmark{4}, H. Furusawa\altaffilmark{4},
T. Hattori\altaffilmark{2}, 
M. Iye\altaffilmark{4}, K. S. Kawabata\altaffilmark{5}, 
N. Kobayashi\altaffilmark{6}, Y. Komiyama\altaffilmark{4}, 
G. Kosugi\altaffilmark{4}, Y. Minowa\altaffilmark{2}, 
Y. Mizumoto\altaffilmark{4}, Y. Niino\altaffilmark{1}, 
K. Nomoto\altaffilmark{7}, 
J. Noumaru\altaffilmark{2}, R. Ogasawara\altaffilmark{4}, 
T.-S. Pyo\altaffilmark{2}, T. Sakamoto\altaffilmark{8}, 
K. Sekiguchi\altaffilmark{4}, Y. Shirasaki\altaffilmark{4},
M. Suzuki\altaffilmark{9}, A. Tajitsu\altaffilmark{2},
T. Takata\altaffilmark{4}, T. Tamagawa\altaffilmark{10,11},
H. Terada\altaffilmark{2}, T. Totani\altaffilmark{1},
J. Watanabe\altaffilmark{4}, T. Yamada\altaffilmark{12}, 
and A. Yoshida\altaffilmark{13}}

\altaffiltext{*}{Based on data collected at Subaru Telescope, which is operated by the National Astronomical Observatory of Japan.}
\altaffiltext{1}{Department of Astronomy, Kyoto University, Kyoto 606-8502, Japan}
\altaffiltext{2}{Subaru Telescope, National Astronomical Observatory of Japan, 650 North A'ohoku Place, Hilo, HI 96720, USA}
\altaffiltext{3}{Department of Physics, Tokyo Institute of Technology, 2-12-1 Ookayama, Meguro-ku, Tokyo 152-8551, Japan}
\altaffiltext{4}{National Astronomical Observatory of Japan, 2-21-1 Osawa, Mitaka, Tokyo 181-8588, Japan}
\altaffiltext{5}{Hiroshima Astrophysical Science Center, Hiroshima University, Higashi-Hiroshima, Hiroshima 739-8526, Japan}
\altaffiltext{6}{Institute of Astronomy, University of Tokyo, 2-21-1 Osawa, Mitaka, Tokyo 181-0015, Japan}
\altaffiltext{7}{Institute for the Physics and Mathematics of the Universe (IPMU), University of Tokyo, 5-1-5 Kashiwanoha, Kashiwa, Chiba 277-8568, Japan}
\altaffiltext{8}{NASA Goddard Space Flight Center, Greenbelt, MD 20771, USA}
\altaffiltext{9}{ISS Science Project Office, ISAS, JAXA, 2-1-1 Sengen, Tsukuba, Ibaraki 305-8505, Japan}
\altaffiltext{10}{Department of Physics, Tokyo University of Science, 1-3 Kagurazaka, Shinjuku-ku, Tokyo 162-8601, Japan}
\altaffiltext{11}{Cosmic Radiation Laboratory, Institute of Physical and Chemical Research, 2-1 Hirosawa, Wako, Saitama 351-0198, Japan}
\altaffiltext{12}{Astronomical Institute, Tohoku University, Sendai 980-8578, Japan}
\altaffiltext{13}{Department of Physics, Aoyama Gakuin University, Sagamihara, Kanagawa 229-8558, Japan}

\begin{abstract}
We present optical and near infrared observations of $Swift$ GRB 080325 
classified as a \lq\lq Dark GRB\rq\rq. Near-infrared observations with 
Subaru/MOIRCS provided a clear detection of afterglow in $K_{s}$ band, 
although no optical counterpart was reported. 
The flux ratio of rest-wavelength optical to X-ray bands of the afterglow 
indicates that the dust extinction along the line of sight to the afterglow 
is $A_{V}$ = $2.7 - 10$ mag. 
This large extinction is probably the major reason for optical faintness 
of GRB 080325. 
The $J - K_{s}$ color of the host galaxy, ($J - K_{s}$ = 1.3 in AB magnitude), 
is significantly redder than those for typical GRB hosts previously 
identified.
In addition to $J$ and $K_{s}$ bands, optical images in $B$, $R_{c}$, $i$', and $z$' 
bands with Subaru/Suprime-Cam were obtained at about one year after the burst, 
and a photometric redshift of the host is estimated to be 
$z_{photo} =$ 1.9. 
The host luminosity is comparable to 
$L^{*}$ at $z \sim 2$ in contrast to the sub-$L^{*}$ property of 
typical GRB hosts at lower redshifts.
The best-fit stellar population synthesis 
model for the host shows that a large dust extinction ($A_{V} = 0.8$ mag) attributes 
to the red nature of the host and that the host galaxy is massive 
($M_{*} = 7.0 \times 10^{10} M_{\odot}$) which is 
one of the most massive GRB hosts previously identified. 
By assuming that the 
mass-metallicity relation for star-forming galaxies at $z \sim 2$ 
is applicable for the GRB host, this large stellar mass 
suggests the high metallicity environment around GRB 080325, 
consistent with inferred large extinction. 
\end{abstract}
\keywords{galaxies: photometry --- gamma rays: burst}

\section{INTRODUCTION}
The origin of a \lq\lq dark\rq\rq gamma-ray burst (GRB) remains one of the most 
serious mysteries of long GRB (hereafter just referred as GRBs) phenomena. 
A dark GRB is characterized by 
the faintness of its optical afterglow. While the X-ray afterglow is 
currently well-explored since the launch of $Swift$, an optical 
and/or near-infrared detection is reported in only about half of cases, 
suggesting the significant fraction of dark events. This optically dark nature 
may be attributed to the (i) large dust extinction around a GRB, (ii) 
attenuation by the neutral hydrogen in the host and/or intergalactic 
medium due to a high redshift event, (iii) intrinsically low 
luminous event of a small GRB fluence, and (iv) low-density medium 
surrounding a GRB progenitor. However, the actual nature of dark GRBs 
has not yet been revealed well.

Recently, the metallicity environment of dark GRBs is getting a lot more 
attention. The low metallicity has been theoretically suggested to be required 
to produce the rapid rotating progenitors and relativistic explosions associated with GRBs 
\citep{2006ARA&A..44..507W,2006A&A...460..199Y}. As for observational studies, 
\cite{2006Natur.441..463F} showed that GRBs preferentially occur in small faint irregular 
galaxies. This suggests the low-metallicity environment around GRBs because 
mass (or luminosity)-metallicity relation exists among galaxies. In fact, 
the low-metallicity nature of GRB hosts based on spectroscopic observations are 
reported by \cite{2006AcA....56..333S}. 
\cite{2008AJ....135.1136M} compared the chemical abundances at the sites of 12 
nearby ($z < 0.14$) type Ic supernovae (SN Ic) that showed broad lines, but had no 
observed gamma-ray burst (GRB), with the chemical abundances in five nearby 
($z < 0.25$) galaxies at the sites of GRBs. 
They showed 
that there is a critical metallicity below which GRBs occur 
in contrast to type Ic SNe, which occur in more metal rich environment. 
In this context, \cite{2009ApJ...702..377K} 
predicted the critical stellar mass of GRB hosts, which is a stellar-mass 
counterpart of the critical metallicity, as a function of redshift, 
assuming an evolution of the mass-metallicity 
relation by \cite{2005ApJ...635..260S} and the presence of the critical 
metallicity \citep{2008AJ....135.1136M}. They showed that most GRB hosts 
collected by \cite{2009ApJ...691..182S} have stellar masses smaller 
than or comparable to the critical mass. 
There are, however, exceptionally massive hosts at $z \gtrsim 0.4$, i.e., possibly 
exceptionally high metallicity: hosts of GRB 020127 
\citep[$z \sim 2$:][]{2007ApJ...660..504B}, GRB 020819, 
and GRB 051022 \citep[$z =$ 0.4 and 0.8, respectively:][]{2009ApJ...691..182S}. 
For GRB 020819, \cite{2010ApJ...712L..26L} reported metallicity larger than the 
critical metallicity at the GRB position as well as in the host. In addition, for GRB 051022 
metallicity of the host is slightly larger than the critical value above mentioned 
\citep{2009AIPC.1133..269G}. 
Both GRBs are classified as dark\footnotemark[1].  
\footnotetext[1]{It is not clear whether GRB 020127 is classified as \lq\lq dark\rq\rq or not 
because of no strong constraint on optical-to-X-ray spectral index of the afterglow 
\citep[$\beta_{\rm OX} < 1.24$:][]{2004ApJ...617L..21J}.}
This suggests that high metallicity environment around a GRB may be 
somehow related to its \lq\lq dark\rq\rq nature. Therefore, investigating the properties of 
massive dark GRB hosts is a key to reveal the origin of dark GRBs.  


The $Swift$ Burst Alert Telescope (BAT) detected the long-duration 
GRB 080325 (T$_{90} = 128.4 \pm 34.2$ s) on 2008 March 25 at 
04:10:32 UT with an initial localization of 3$\arcmin$ radius 
\citep{2008GCN..7512....1V}.
The $Swift$ X-Ray Telescope (XRT) 
began observing the field at 151.9 s after the BAT trigger, 
and found a bright, uncatalogued X-ray source 
\citep{2008GCN..7512....1V}. The enhanced XRT position was reported 
by \cite{2008GCN..7513....1O} by using 199 s of overlapping XRT Photon 
Counting mode 
with an uncertainty of 2$\arcsec$.6 radius. After the trigger, several 
attempts were made to identify the afterglow in optical and near-infrared 
wavelength 
\citep{2008GCN..7516....1D,2008GCN..7518....1B,
2008GCN..7520....1C,2008GCN..7522....1K,
2008GCN..7529....1I,2008GCN..7563....1M}. 
However, no detection of the afterglow or host galaxy has been reported 
in optical and near-infrared wavelength, suggesting that the GRB 080325 
is a dark GRB. 
In this paper we report the successful photometric identification 
of the 
afterglow of GRB 080325 in near-infrared wavelength and its host galaxy at 
$z \sim 2$, and present a case study of properties of the 
afterglow and its host galaxy of a dark GRB.
This paper is organized as follows: in \S 
\ref{NEAR-INFRARED AND OPTICAL IMAGING WITH SUBARU} the 
near-infrared and optical observations and data analysis are presented. 
In \S \ref{GRB 080325 AS A DARK GRB} 
we show the 
faintness of the optical and near-infrared afterglow of GRB 080325 
compared with its X-ray afterglow, and dust extinction along the line of 
sight to GRB 080325. Results of spectral energy distribution fitting 
analysis of the GRB 080325 host and discussion are presented 
in \S \ref{MASSIVE RED HOST GALAXY AT} and \S 
\ref{COMPARISON WITH OTHER GRB AFTERGLOWS AND HOSTS}, respectively. 

Throughout this paper we use the AB magnitude system 
and the cosmological parameters of 
$H_{0}$ = 71 km s$^{-1}$ Mpc$^{-1}$, $\Omega_{M}$ = 0.27, and 
$\Omega_{\Lambda}$ = 0.73.

\section{NEAR-INFRARED AND OPTICAL IMAGING WITH SUBARU}
\label{NEAR-INFRARED AND OPTICAL IMAGING WITH SUBARU}

We observed the GRB 080325 field as a part of a target-of-opportunity 
program with the Subaru telescope. 
We took images of the field using Multi-Object InfraRed 
Camera and Spectrograph (MOIRCS) in $J$ and $K_{s}$ bands at 8.7 
hours after the burst and another set at a day latter. 
A standard dithering method was 
employed during the exposure, with a single exposure of 
50 s $\times$ 2 coadds (first night) and 40 s $\times$ 2 coadds (second night) in $K_{s}$ band, 
and with 90 s (first night) and 150 s (second night) in $J$ band. 
Total exposure times in $J$ and $K_{s}$ bands are 900 s and 3300 s 
for the first night, and 2250 s and 5040 s for the second night, respectively. 
The weather condition was clear for both nights. 
Seeing sizes were roughly 0$\arcsec$.6 in $K_{s}$ band 
and 0$\arcsec$.8 in $J$ band. 
A standard star field (FS27) was also observed during the first night at
the end of the observation. 
The data reduction was performed with the MCSRED package (ver.20080317) 
using all-in-one task (mcsall). 

\begin{figure}
\begin{center}
\includegraphics[scale=0.3]{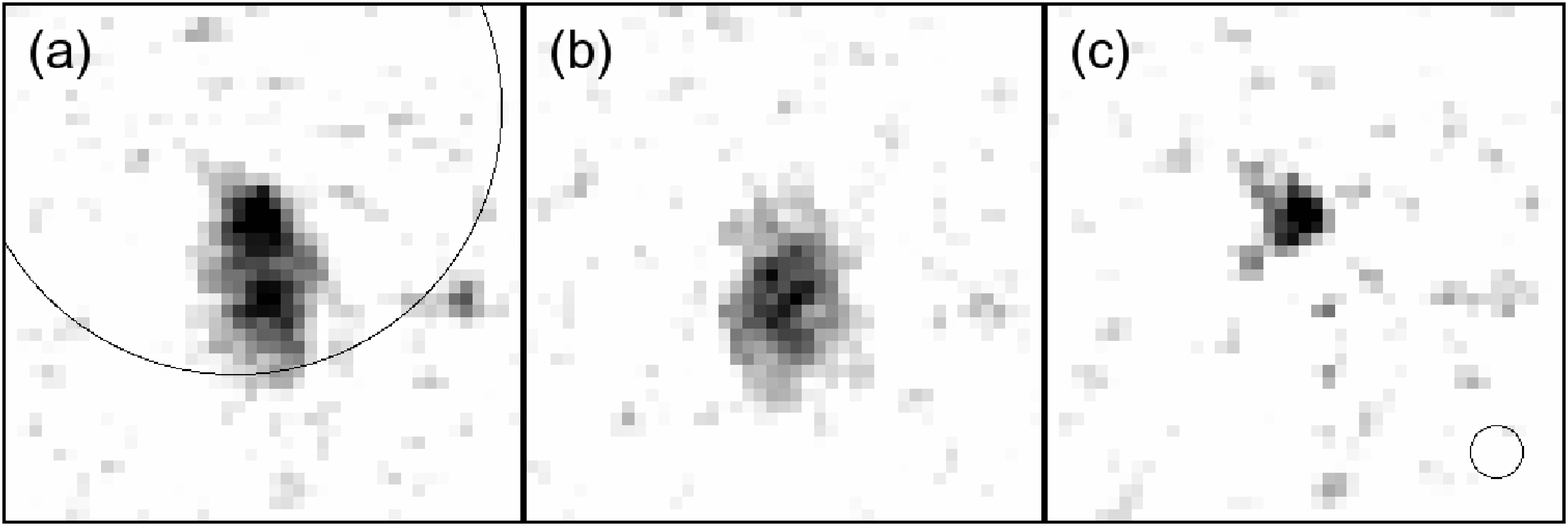}
\caption{
$K_{s}$ band images (5$\arcsec$.0 $\times$ 5$\arcsec$.0) of GRB 080325 host galaxy at 8.7 hours 
(a), and 33.5 hours (b) after the burst. (c) Subtracted image. 
North is up and east is to left. Large and small circles represent 
the enhanced $Swift$ XRT error circle \citep{2008GCN..7513....1O} and a seeing size of 
$\sim$ 0$\arcsec$.5 $\phi$, respectively. 
}
\label{figure1} 
\end{center}
\end{figure}

We detected a faint extended object as well as a point-like spot 
at the northern edge of it 
in the enhanced $Swift$ XRT error circle \citep{2008GCN..7513....1O} in the 
first-night $K_{s}$-band image as shown in Figure \ref{figure1} (a). In the first-night 
$J$-band image, although the faint extended object was seen, the northern source 
was not detected. In the second-night $K_{s}$-band image, the north spot was not 
detected significantly as shown in Figure \ref{figure1} (b). By subtracting the second image 
from the first image, the point-like spot can be isolated clearly 
(Figure \ref{figure1}(c)); a total magnitude (the spot and the extended object) decreased 
by 0.2 mag. No significant variation was detected in $J$ band. 
Based on the positional coincidence, 
the FWHM of the northern spot comparable to a seeing size, and 
its fading behavior in $K_{s}$-band, 
we conclude that the spot in the north is the afterglow. 
The coordinates of the afterglow are $\alpha$ = $18^{h}31^{m}34^{s}.23$ and 
$\delta$ = $+36^{\circ}31^{\prime}24^{\prime \prime}.8$ (J2000, uncertainty 0$\arcsec$.2). 
The aperture (1$\arcsec$.2 $\phi$) magnitudes of 
the afterglow at 8.7 hours after the burst are $J > 22.2$ mag and $K_{s} = 23.4$  $\pm$ 0.18 mag. 
We also conclude that the faint extended source just south of the afterglow is the host galaxy of 
GRB 080325. 
The aperture ($2\arcsec.0$ $\phi$) magnitudes of the host galaxy are 
$J = 23.0 \pm 0.18$ mag and $K_{s} = 21.7 \pm 0.06$ mag. These magnitudes 
substantially represent total magnitudes because of convergence of growth 
curves at the aperture size of 2$\arcsec$.0 in $J$ and $K_{s}$ bands. 
The offset distance of the afterglow from the galaxy center 
($\delta r = 0\arcsec.8 = 7$ kpc) is large among other long GRBs, although 
the offset distance normalized by the half light radius of the 
host ($\delta r/r_{e}$ = 1.3) is a typical value \citep{2002AJ....123.1111B,2010ApJ...708....9F}.
It should be mentioned here that probabilities of a chance coincidence of a foreground/background 
galaxy are small; 
the expected numbers of galaxies brighter than the magnitudes 
within $r \sim 1\arcsec.0$ from the position of the afterglow 
are $\sim 0.03 - 0.04$ in optical bands and $\sim 0.01$ in $K_{s}$-band 
\citep{2001MNRAS.323..795M,2009A&A...507..195R,2003ApJ...595...71C}.

\begin{figure}
\begin{center}
\includegraphics[scale=0.15]{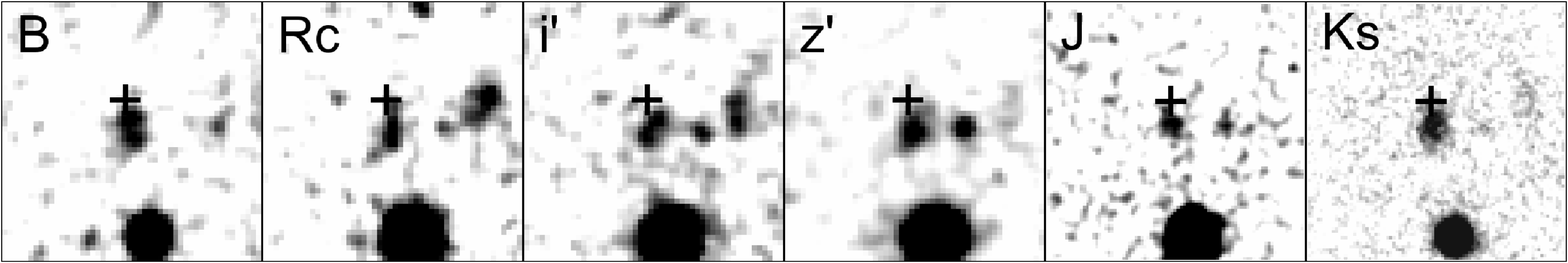}
\caption{
Optical-to-near-infrared images (10$\arcsec$.0 $\times$ 10$\arcsec$.0) of the GRB host galaxy 
obtained by Subaru/Suprime-Cam and MOIRCS. North is up and east is to left. 
Cross shows the position of the GRB 080325 afterglow detected in $K_{s}$ band.
}
\label{figure2}
\end{center}
\end{figure}

We obtained optical images of the GRB 080325 host with 
Subaru/Suprime-Cam in $B$, $R_{c}$, $i$', and $z$' bands on 2009 April 22 
($B$, $R_{c}$, and $z$') and 2009 July 17 UT ($i$') 
to reveal the host properties. 
The total exposure times for $B$, $R_{c}$, $i$', and $z$' bands are 900 s, 360 s, 
360 s, and 1200 s, respectively. The weather condition was photometric and a 
typical seeing size was about 0$\arcsec$.6 nearly identical to 
that of near-infrared images.
The standard data reduction was performed 
for all images, i.e., bias subtraction, flat fielding, distortion correction, 
and sky subtraction using SDFRED package constructed by 
\cite{2002AJ....123...66Y} and \cite{2004ApJ...611..660O}. 
The images were calibrated with observations of SA111 standard 
stars \citep{2009AJ....137.4186L}. 
Obtained images are shown in Figure \ref{figure2}.
We successfully detected the host galaxy with $B$ = 25.7 $\pm$ 0.11 mag, 
$R_{c}$ = 25.5 $\pm$ 0.16 mag, $i'$ = 24.9 $\pm$ 0.16 mag, and $z'$ = 24.5 $\pm$ 0.07 mag 
derived by aperture photometry using a 2$\arcsec$.0 aperture size. 
The magnitude error includes a photometric error of the host galaxy 
calculated by IRAF phot task and a systematic error of a zero point 
in each band. All magnitudes described above are corrected for 
the foreground extinction by Milky Way \citep{1998ApJ...500..525S}. 

\section{GRB 080325 AS A DARK GRB}
\label{GRB 080325 AS A DARK GRB}
\begin{figure}
\begin{center}
\includegraphics[scale=0.7,angle=-90]{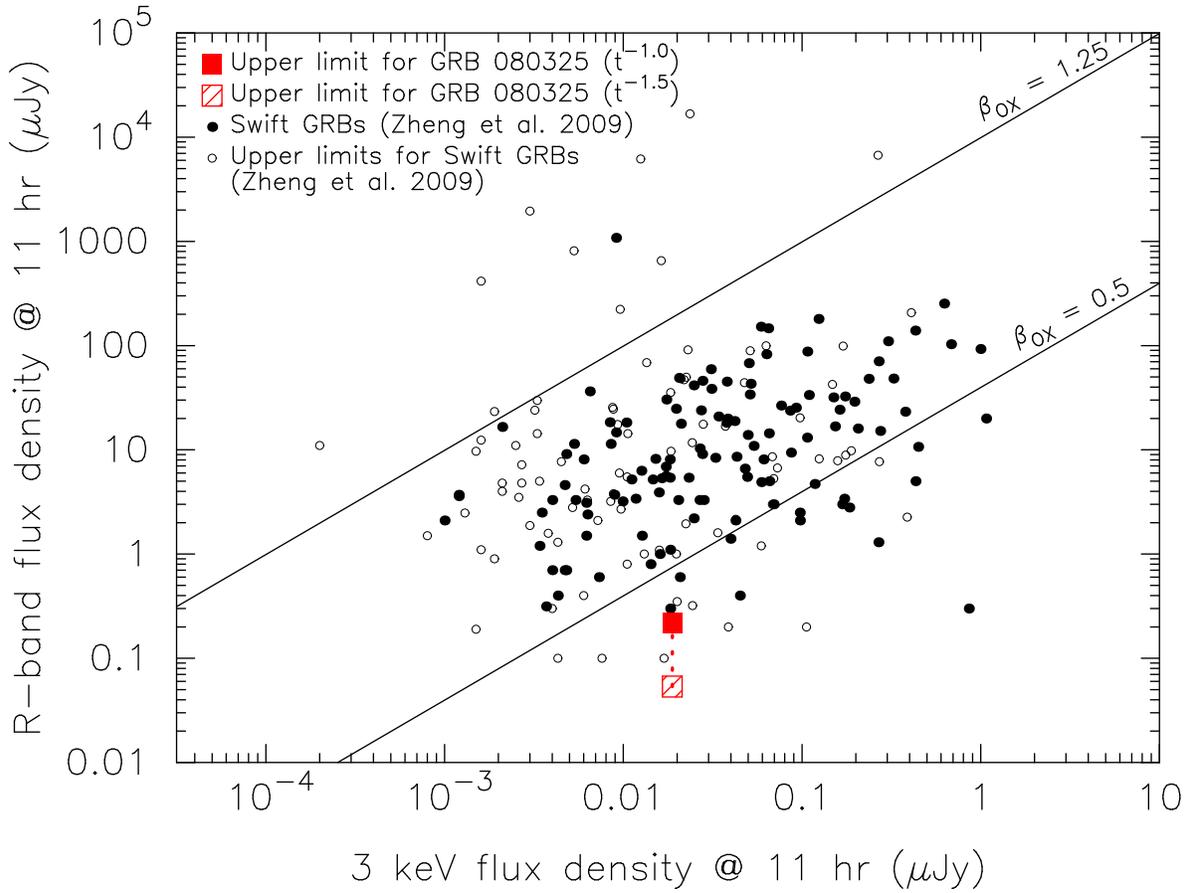}
\caption{
$R$-band flux density versus 3 keV X-ray flux density at 11 hours 
after the burst 
for long GRBs \citep{2009RAA.....9.1103Z}. Filled and 
open circles refer to optical detections reported and optical upper 
limits, respectively. Two solid lines show the theoretical 
limiting values of $\beta_{\rm OX}$ = 0.5 and 1.25. Estimated 
two upper limits on $R$-band flux densities of the GRB 080325 
afterglow are plotted by red filled ($t^{-1.0}$) and hatched ($t^{-1.5}$) 
squares connected by a dotted line.
}
\label{figure3}
\end{center}
\end{figure}

No optical and near-infrared detection of the 
afterglow has been reported other than our detection 
\citep{2008GCN..7524....1T}. 
This may indicate that GRB 080325 is a dark GRB. 
The definition of dark GRBs based only on the optical faintness of afterglow, 
however, is relatively unphysical since the optical detectability of 
an afterglow strongly depends on instruments, weather conditions, and 
starting time of follow-up observations. Thus 
\cite{2004ApJ...617L..21J} proposed a criterion based on an optical-to-X-ray 
spectral index at 11 hours after the burst, i.e., $\beta_{\rm OX}$ = 
log \{$f_{\nu}$(R)/$f_{\nu}$(3 keV)\}/log ($\nu_{3\rm keV}/\nu_{\rm R}$).
If $\beta_{\rm OX}$ of an afterglow 
is below 0.5, the GRB is defined as \lq\lq dark\rq\rq.
In the simplest fireball models which have been successfully 
used to interpret the observed properties of GRB afterglows 
\citep{1997MNRAS.288L..51W, 1998ApJ...497L..17S}, $\beta_{\rm OX}$ is expected 
to be $0.5 < \beta_{\rm OX} < 1.25$. In fact, most of afterglows show 
$\beta_{\rm OX}$ between 0.5 to 1.25 as seen in Figure \ref{figure3}. 
The $\beta_{\rm OX}$ index would be little affected by 
the density of the GRB environment and the GRB luminosity. 
This is because these two cases 
tend to uniformly scale an overall spectrum of an afterglow, 
resulting in a nearly unchanged $\beta_{\rm OX}$ value. 
Based on the X-ray light-curve and spectral analyses of the GRB 080325 
reported in $Swift$/XRT GRB light curve\footnotemark[2] and 
spectral\footnotemark[3] repository, 
\footnotetext[2]{http://www.swift.ac.uk/xrt\_curves/}
\footnotetext[3]{http://www.swift.ac.uk/xrt\_spectra/} 
the X-ray flux in a $0.3 - 10$ keV band at 11 hours was converted to the 
flux density at 3 keV using the measured X-ray spectral index of 
$\beta _{\rm X}$. To estimate the $R$-band flux density of the afterglow at 11 
hours, we assumed two possible 
cases of temporal declining index ($f_{\nu} \propto t^{-1.0}$ and $t^{-1.5}$) 
indicated by X-ray light curve of the afterglow \citep{2007A&A...469..379E}. 
These indices well agree with a typical index of other X-ray afterglows.
By using these temporal indices, the $I$-band upper limit measured with the 
1.5m telescope 
at Sierra Nevada Observatory at 39.46 minutes after the burst 
\citep{2008GCN..7516....1D} 
is extrapolated to two upper limits at 11 hours. These $I$-band upper limits 
are extrapolated to $R$-band upper limits assuming a typical optical spectral 
index of $\beta_{\rm O/NIR}$ = 1.0. As a result, two upper limits on 
$\beta_{\rm OX}$ is estimated to be 0.14 and 0.33 (Figure \ref{figure3}). 
We also confirmed the power-law interpolations between $I$-band upper limits 
and $K_{s}$ band flux densities at 11 hours result in $\beta_{\rm OX}$ 
slightly smaller than these values, 
assuming above two temporal declining indices 
(the power-law indices in this interpolation are 
$\beta _{\rm O/NIR}$ = 1.6 and 2.9 for $t^{-1}$ and $t^{-1.5}$). 
Therefore, GRB 080325 is classified as a dark GRB.

\begin{figure}
\begin{center}
\includegraphics[scale=0.7]{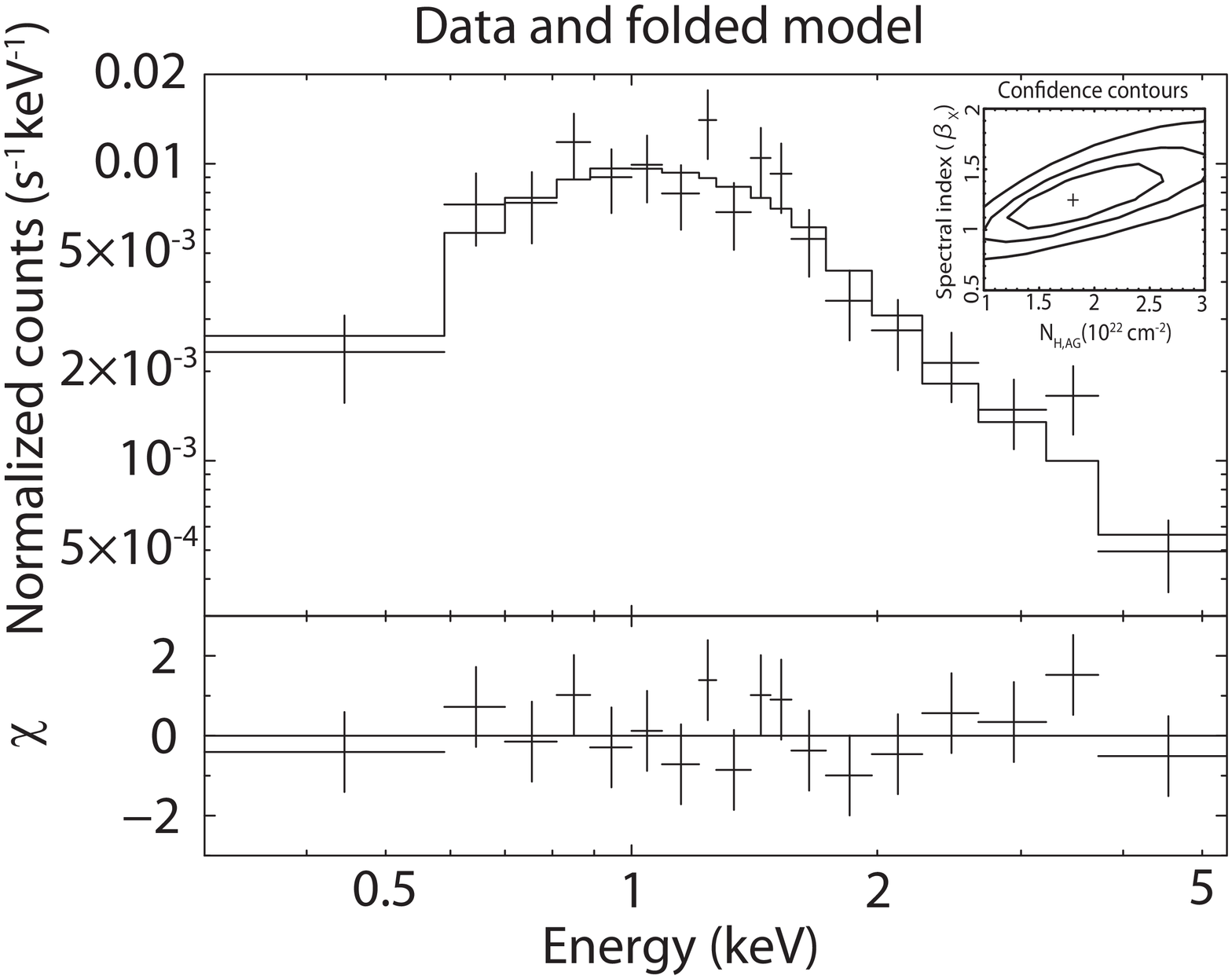}
\caption{
(Top) Observed spectrum of the X-ray afterglow averaged from 
1 to 16 hours after the burst and the best-fit spectral model (solid line). 
Inserted three contours indicate 68\%, 90\%, and 99\% confidence levels in 
the parameter space of $\beta_{\rm X}$ and N$_{\rm H,AG}$, and 
a central cross is best-fit result. 
(Bottom) Delta chi-square at each energy bin.
}
\label{figure4}
\end{center}
\end{figure}

Given a typical optical spectral index of $\beta _{\rm O/NIR}= 1.0$ 
and the observed $K_{s}$ magnitude of the afterglow, the expected $J$-band magnitude 
of the afterglow at the first night is estimated to be $J = 24.0$ mag, 
which is fainter than the $J$-band limiting magnitude of the first-night image. 
This result is unchanged even if two power-law indices of 
$\beta _{\rm O/NIR}$ = 1.6 and 2.9 mentioned above are assumed.
Therefore, it is unclear whether large attenuation by dust or intergalactic 
medium are present or not from the $J - K_{s}$ color of the afterglow. 
Another method to examine the attenuation along the line of sight to the GRB is a 
comparison between flux densities of the X-ray and optical afterglows. 
As mentioned above, \lq\lq intrinsic\rq\rq (rest-wavelength obscuration-free) 
optical-to-X-ray spectral index ($\beta_{\rm OX,int}$) 
is expected to be 0.5 - 1.25. If we estimate the \lq\lq intrinsic\rq\rq 3 keV flux density 
of the afterglow ($f_{3keV,int}$), a possible range of the \lq\lq intrinsic\rq\rq $R$-band 
flux density ($f_{R,int}$) can be obtained. Thus, the comparison between $f_{R,int}$ and 
observed rest-wavelength $R$-band flux density of the afterglow ($f_{R,obs}$) 
enables us to estimate an extinction along the line of sight to the GRB.
So as to estimate a possible range of $f_{R,int}$, 
we obtained the X-ray spectrum of the GRB 080325 
afterglow averaged from 1 to 16 hours after the burst 
from the Swift/XRT GRB spectral repository, 
and performed spectral fitting analysis as shown in Figure \ref{figure4}.
The X-ray spectral model includes a fixed extinction by Milky Way 
(N$_{\rm H}$ = $3.8 \times 10^{20}$ cm$^{-2}$), an extinction to the 
afterglow of GRB 080325 at $z = 1.9$ (N$_{\rm H,AG}$), 
and a power-law X-ray spectral index of $\beta_{\rm X}$. The redshift 
of the afterglow assumed here is based on 
spectral energy distribution (SED) fitting analysis 
for the host galaxy described in section 
\ref{MASSIVE RED HOST GALAXY AT}. 
Using the derived best-fit parameters of 
N$_{\rm H,AG}$ = $(1.8^{+0.8}_{-0.6}) \times 10^{22}$ cm$^{-2}$ and 
$\beta_{\rm X}$ = 1.2$^{+0.3}_{-0.2}$, 
the X-ray flux in a $0.3 - 10$ keV band at 11 hours extracted from 
X-ray light curve was converted to $f_{3keV,int} = 0.20$ $\mu$Jy. 
Based on $f_{3keV,int}$ and $\beta_{\rm OX,int}$, 
we calculated a possible range of $f_{R,int}$. 
With the Calzetti extinction law \citep{2000ApJ...533..682C}, the extinction 
along the line of sight to the GRB is estimated to be 
$2.7$ mag $< A_{V} < 10$ mag. 
If extinction laws for Milky Way \citep{1992ApJ...395..130P} and SMC 
\citep{1984A&A...132..389P,1985A&A...149..330B} 
are assumed, the extinction is $2.8$ mag $< A_{V} < 11$ mag 
for both cases. 
Since these values are corresponding to 
$\beta_{\rm OX,int}$ = 0.5 and 1.25, respectively, we can not 
exclude the extremely strong dust extinction of $A_{V} > 10$ mag 
if $\beta_{\rm OX,int} > 1.25$ is possible, e.g., there is one exceptional 
object with $\beta_{\rm OX} > 1.25$ as shown in Figure \ref{figure3}. 
As an alternative, the best-fit result of 
N$_{\rm H,AG}$ = $1.8 \times 10^{22}$ cm$^{-2}$ can be converted to 
$A_{V} \sim 10$ assuming a typical ratio of $A_{V}/N_{\rm H}$ for the 
Local Group galaxies \citep[Milky Way, SMC, and LMC:][]{2007MNRAS.377..273S}.
In any cases $A_{V}$ along the line of sight to the GRB is 
larger than $A_{V}$ for the entire host derived below. 

\section{MASSIVE RED HOST GALAXY AT $z \sim 2$}
\label{MASSIVE RED HOST GALAXY AT}

In order to estimate a redshift and to reveal host properties we 
made SED fitting for the host. 
The stellar properties of the GRB 080325 host are examined using 
the observed multi-band photometry and stellar population synthesis models 
including emission lines typical for local star-forming galaxies 
(P$\acute{\rm E}$GASE.2) constructed by \cite{1997A&A...326..950F} and 
\cite{1999astro.ph.12179F}. Assumed star-formation histories (SFHs) include 
(i) instantaneous burst, (ii) exponentially declining star-formation rate 
(SFR), i.e., $e^{-\frac{t}{\tau}}$ with $\tau$ = 1 Myr, 10 Myr, 100 Myr, 
1 Gyr, and 10 Gyr, and (iii) 
constant SFR. The initial mass function (IMF) used here is 
Salpeter IMF \citep{1955ApJ...121..161S}, and the extinction 
law is adopted from \cite{2000ApJ...533..682C}. 
Although we also examined Milky Way and SMC extinction laws, these 
provided very poor fitting results. 
We used {\it SEDfit} software package (M. Sawicki 2010, in preparation) 
which is essentially the same as that by \cite{1998AJ....115.1329S}.

\begin{figure}
\begin{center}
\includegraphics[scale=1.2,angle=-90]{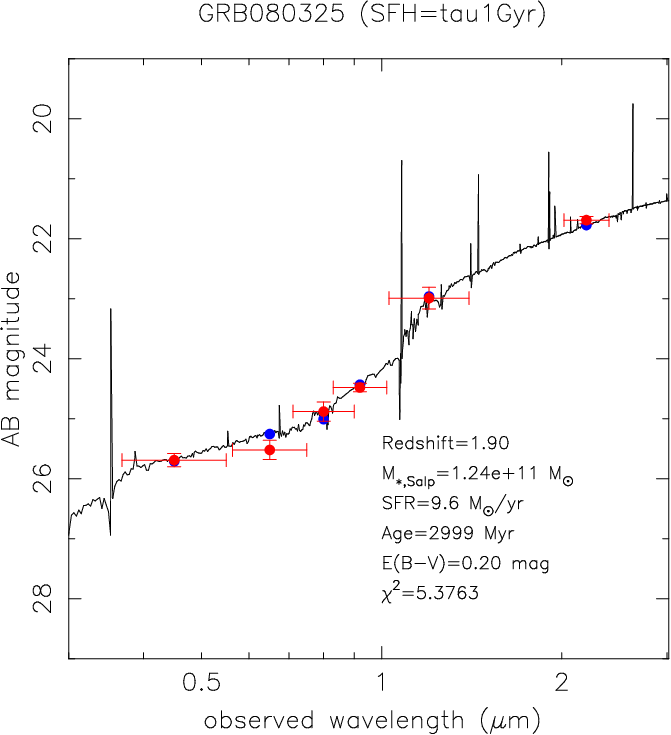}
\caption{
Observed SED of the GRB 080325 host galaxy (red dots with error bars) 
and the best-fit synthetic spectrum (black solid line) with P$\acute{\rm E}$GASE.2, 
which is a code to compute the galaxy stellar population synthesis models including 
emission lines from ionized gas \citep{1997A&A...326..950F,1999astro.ph.12179F}. 
Blue dots without error bars show AB magnitude of the best-fit model in 
observed bands. The best-fit model shows the dusty ($E(B-V) = 0.2$ mag) 
massive ($M_{*,\rm Salp} = 1.24 \times 10^{11} M_{\odot}$) 
starforming (SFR = $9.6 M_{\odot}$ yr$^{-1}$) host galaxy at $z = 1.9$. 
}
\label{figure5}
\end{center}
\end{figure}

\begin{figure}
\begin{center}
\includegraphics[scale=0.7,angle=-90]{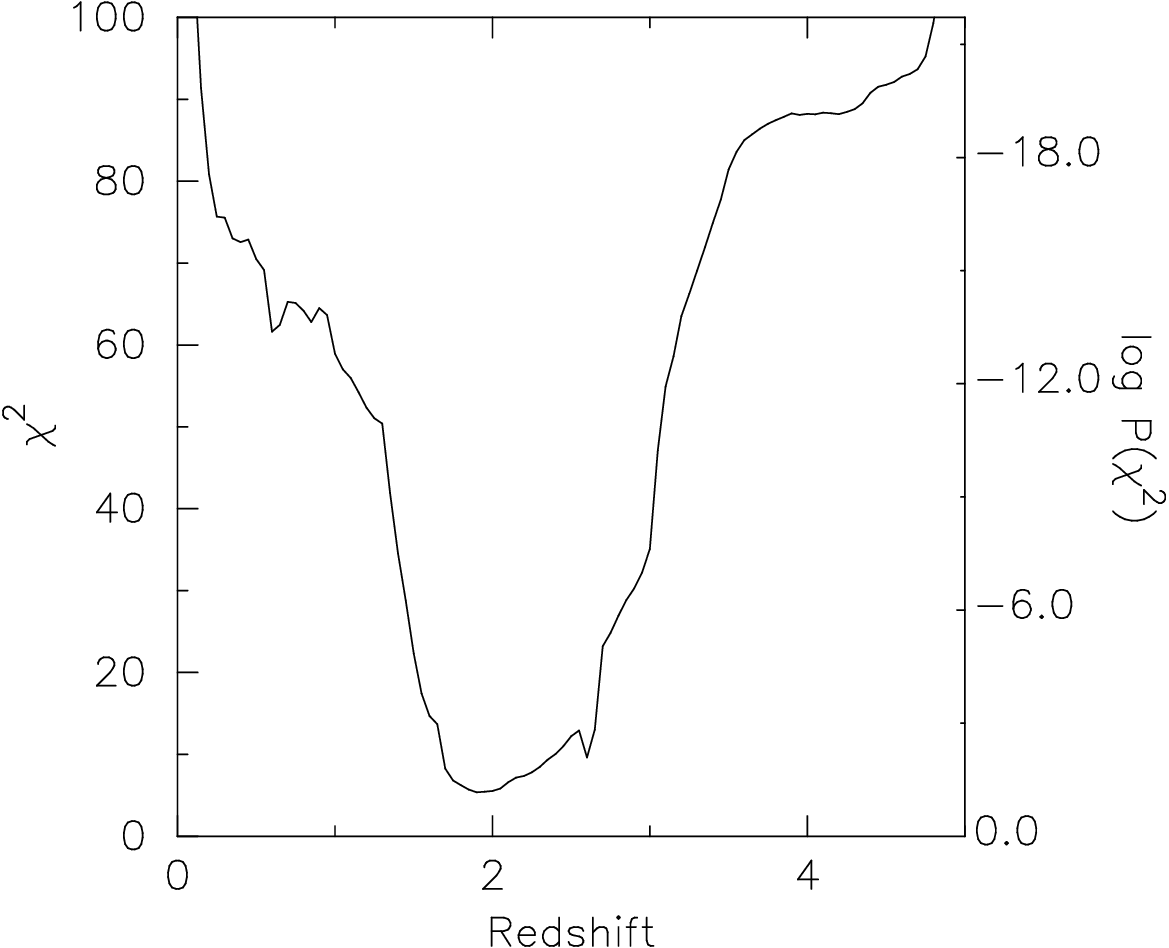} 
\caption{
$\chi^{2}$ (left-side $y-$axis) or logarithm of probability (right-side $y-$axis) 
as a function of redshift for the GRB 080325 host galaxy using stellar population 
synthesis models including emission lines from ionized gas (P$\acute{\rm E}$GASE.2) 
constructed by \cite{1997A&A...326..950F} and \cite{1999astro.ph.12179F}. 
A step size of redshift is $\Delta z$ = 0.05.
At each redshift, a minimum value of $\chi^{2}$ is adopted among 
instantaneous burst, exponentially declining 
($\tau$ = 1 Myr, 10 Myr, 100 Myr, 1 Gyr, and 10 Gyr), and constant SFR 
models. The best-fit redshift is $z = 1.9$ for $\tau$ = 1 Gyr model. 
}
\label{figure6}
\end{center}
\end{figure}

The SFH of the best-fit stellar population synthesis model is the 
$\tau$ (= 1 Gyr) declining model ($\chi^{2}=5.4$),  
and the best-fit result and output parameters 
of the host galaxy are summarized in Figure \ref{figure5}. 
However $\tau$ = 10 Gyr and constant SFR models can also reasonably 
reproduce the observed SED ($\chi^{2}=7.8$). These models give that the host is at 
$z = 1.8 - 2.2$ with the best estimate of 1.9 (see Figure \ref{figure6}), 
and the dust extinction is large ($A_{V} = 0.8 - 1.6$ mag, $E(B-V)=0.2 - 0.4$ mag). 
The $M_{*,\rm Salp}$ (stellar mass for Salpeter IMF), SFR, 
and age of the host derived from SED fitting are 
$(1.2 - 1.8) \times 10^{11} M_{\odot}$, 
$10 - 80 M_{\odot}$ yr$^{-1}$, and $3.0 - 3.5$ Gyr, although 
uncertainties of the SFR and age are relatively large. 

\begin{figure}
\begin{center}
\includegraphics[scale=0.7,angle=-90]{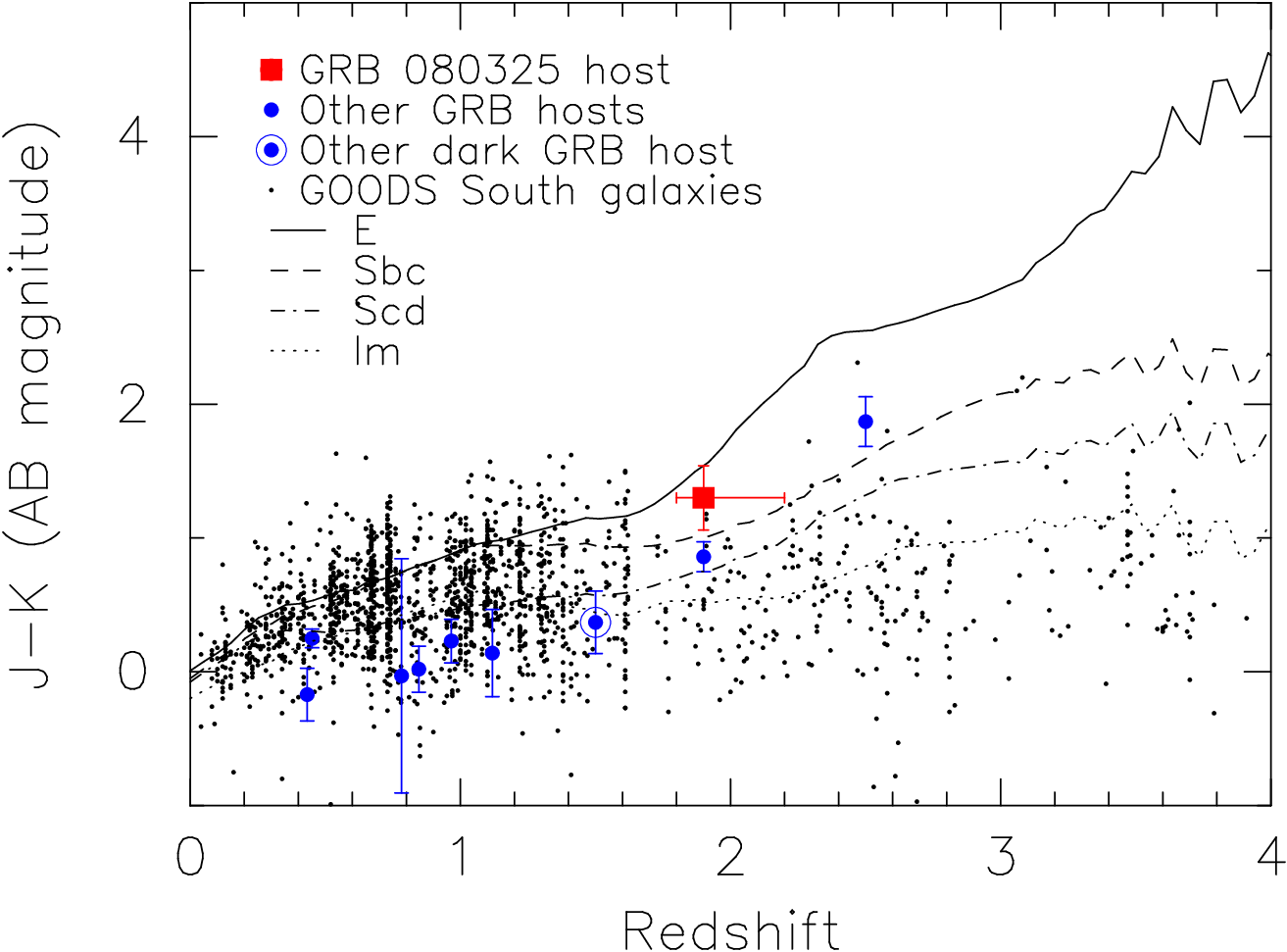}
\caption{
$J - K_{s}$ apparent color in AB magnitude as a function of redshift, 
for GRB hosts  
\citep[large blue dots with error bars:][]{2006ApJ...647..471L,2007ApJ...660..504B, 
2008ApJ...681..453J,2009ApJ...691..182S}
and GOODS South galaxies with spectroscopic redshifts 
\citep[small dots:][]{2006A&A...449..951G}. 
Circled large blue dot shows a dark GRB host defined by optical-to-X-ray spectral 
index of $\beta_{\rm OX} < 0.5$ \citep{2009RAA.....9.1103Z}.
Typical templates of galaxies at $z$ = 0 \citep{1980ApJS...43..393C} 
are represented by different line types 
(elliptical, Sbc, Scd, and irregular types are shown by 
solid, dashed, dot-dashed, and dotted lines, respectively). 
Red square corresponds to the GRB 080325 host.
}
\label{figure7}
\end{center} 
\end{figure}

The rest-frame $B$-band absolute 
magnitude is $M_{B}$ = $-21.8$ calculated from the best-fit spectral 
template. This magnitude is comparable to $L^{*}$ at $1.8 < z < 2.0$ 
\citep{2005ApJ...631..126D}, or 3 $L^{*}$ if the $B$-band luminosity is 
corrected for the dust extinction. As shown in Figure \ref{figure7}, 
the observed color of the GRB 080325 host ($J - K_{s}$ = 1.3 mag) 
is significantly redder than any other GRB hosts 
\citep[large blue dots with vertical error bars:][]
{2006ApJ...647..471L,2007ApJ...660..504B,2008ApJ...681..453J,2009ApJ...691..182S}
at $z \le 2$, GOODS South galaxies with known spectroscopic redshifts 
\citep[small black dots:][]{2006A&A...449..951G} and local-galaxy 
templates \citep[different line types:][]{1980ApJS...43..393C} 
redshifted to $z \sim 2$ without galaxy evolution except for an elliptical galaxy. 
These properties of the host galaxy, 
i.e., the luminous dusty (red-color) starforming host, are significantly different 
from those of typical GRB hosts previously studied by 
\cite{2003A&A...400..499L}. We note that these results do not change 
even if population synthesis models without including nebular emission lines 
are examined for the SED fitting analysis. Hereafter, we adopt parameters of the 
host galaxy derived from the best-fit model with emission lines. 

\section{COMPARISON WITH OTHER GRB AFTERGLOWS AND HOSTS}
\label{COMPARISON WITH OTHER GRB AFTERGLOWS AND HOSTS}
\subsection{OPTICAL FAINTNESS DUE TO DUST EXTINCTION}
\label{OPTICAL FAINTNESS DUE TO DUST EXTINCTION}
Recently \cite{2009AJ....138.1690P} have made 
a uniform sample of dark GRBs. Among 29 GRB hosts 
rapidly observed with the Plomar 60-inch telescope from 
$Swift$ burst, they selected 14 dark GRBs based on $\beta_{\rm OX}$ at 1000 s 
after the burst. 
They inferred that a 
significant fraction of dark GRBs occurs in highly obscured regions, 
although the host galaxies of dark GRBs seem to have normal optical 
colors. This suggests that the source of obscuring dust is 
local to the vicinity of the GRB progenitor or highly unevenly 
distributed within the host galaxy. \cite{2009AJ....138.1690P} 
demonstrated that the typical dust extinction along the line of sight to 
dark GRBs is 0.5 mag $< A_{V}$ $<$ 10.0 mag. The dust extinction along the 
line of sight to GRB 080325 is similar to these values. 
However, GRB 080325 host is characteristic in the sense that the large extinction is 
observed both for the entire host and the line of sight to the 
GRB, in contrast to most of other dark GRBs which show relatively weak extinction for 
the whole host galaxies, 
i.e., $A_{V} < 0.5$ mag \citep{2009AJ....138.1690P}. 
Another case of dark GRB with a large extinction over the whole host galaxy, 
is GRB 051022. GRB 051022 
is classified as dark with $\beta_{\rm OX} < -0.05$, whose 
host galaxy shows a relatively large 
dust extinction of $A_{V} \sim  1.0$ mag derived from SED fitting 
analysis \citep{2007ApJ...669.1098R, 2007A&A...475..101C}. 
The amount of the extinction in the line of sight toward the 
GRB ($A_{V} = 4.4$ mag), which is required to suppress the optical 
afterglow to the observed limits, is clearly higher than the $A_{V}$ 
value found from the host SED. 
\cite{2007ApJ...669.1098R} attributed this significant 
discrepancy between two values of the dust extinction to the effect 
of the GRB position in the host where the line of 
sight crosses a dense region in the host. As for GRB 080325, 
the dust extinction along the line of sight to the burst is 
larger than that for the host as a whole and 
the GRB is located at the outskirt of the host galaxy. Therefore, the 
local dusty environment 
such as circumstellar matter and/or molecular cloud 
around GRB 080325 is likely to be a major reason for the optical 
faintness of GRB 080325 relative to X-ray brightness rather than extinction by 
dust distributed over the entire host. 

\subsection{MASSIVE GRB HOST AND ORIGIN OF A DARK GRB}
\begin{figure}
\begin{center}
\includegraphics[scale=0.7,angle=-90]{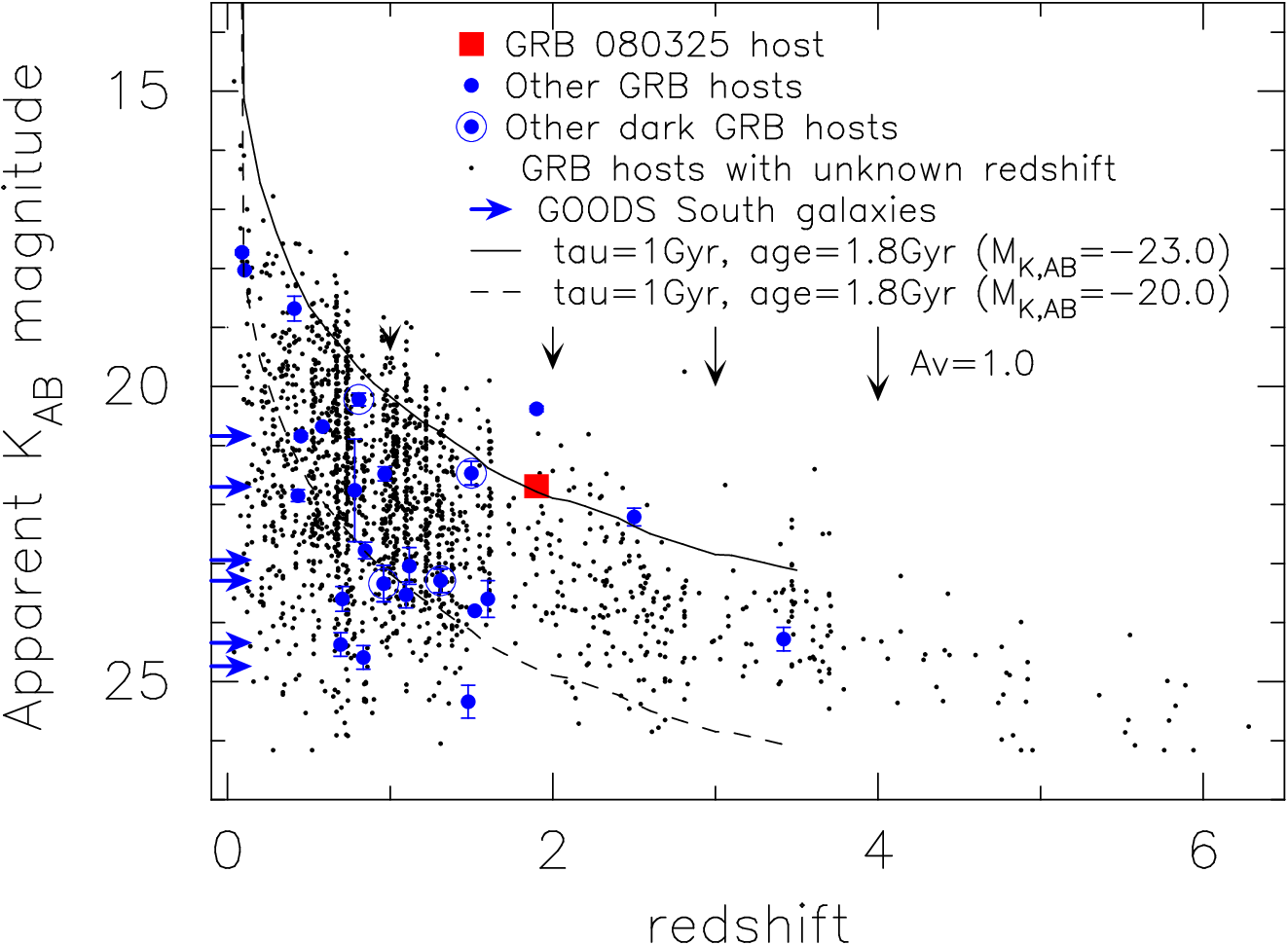}
\caption{
Apparent magnitudes as a function of redshift for GRB hosts 
with and without known redshift 
\citep[large blue dots and arrows:][]{2002ApJ...566..229C,2003A&A...400..499L,
2006ApJ...647..471L,2007ApJ...660..504B,
2008ApJ...681..453J,2008MNRAS.388.1743T, 
2009ApJ...691..182S} 
and GOODS South field galaxies with spectroscopic redshift 
\citep[small black dots:][]{2006A&A...449..951G}. 
Circled large blue dots show dark GRB hosts defined by optical-to-X-ray spectral 
index of $\beta_{\rm OX} < 0.5$ 
\citep{2004ApJ...617L..21J,2009RAA.....9.1103Z,2009ApJ...699.1087V}. 
Black arrows are extinction vectors for 
$A_{V}$ = 1.0 mag at each redshift, given the Calzetti extinction law 
\citep{2000ApJ...533..682C}. Solid and dashed lines show tracks for model galaxies  
($\tau$ declining models) with $M_{K}$ 
= $-23$ and $-20$. GRB 080325 host is shown by a red square.
}
\label{figure8}
\end{center}
\end{figure}

As mentioned in \S \ref{MASSIVE RED HOST GALAXY AT}, 
the host of GRB 080325 is as luminous as $L^{*}$ at $z \sim 2$.
This property is also seen in the apparent 
$K_{s}$-band magnitude as shown in Figure \ref{figure8}. 
The host of GRB 080325 is located in the bright end of the apparent 
$K_{s}$ magnitude distribution of the GOODS south galaxies at the same 
redshift range. It is contrary to the distribution of other GRB hosts, 
which are slightly fainter than that of the GOODS south galaxies. 
In this figure, the host of GRB 080325 is denoted by a red square. 
Previously identified GRB hosts with and without 
known redshift are shown by large blue dots and arrows 
\citep{2002ApJ...566..229C, 2003A&A...400..499L, 
2006ApJ...647..471L, 2007ApJ...660..504B, 2008ApJ...681..453J, 
2008MNRAS.388.1743T, 2009ApJ...691..182S} and GOODS South field 
galaxies with spectroscopic redshifts are indicated by 
small black dots \citep{2006A&A...449..951G}. Black arrows 
represent extinction vectors for $A_{V}$ = 1.0 mag at each redshift 
assuming the Calzetti extinction law \citep{2000ApJ...533..682C}. 
For comparison, tracks of model galaxies 
($\tau$ declining models used in \S \ref{MASSIVE RED HOST GALAXY AT}) 
with $M_{K} = -23$ mag and $-20$ mag 
without extinction are shown by 
the solid and dashed lines. 

\begin{figure}
\begin{center}
\includegraphics[scale=0.7,angle=-90]{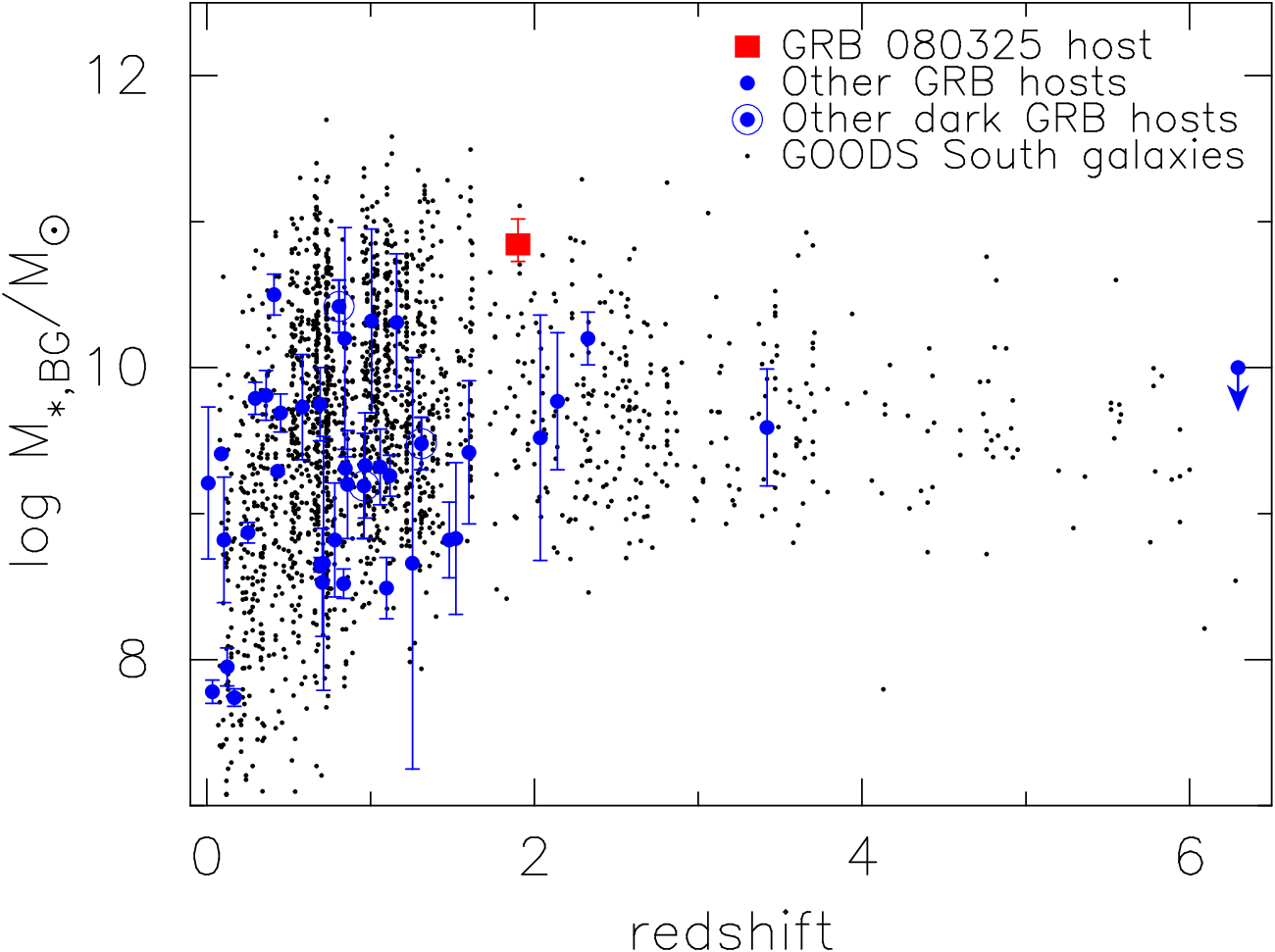}
\caption{
Stellar masses derived from the best-fit SED model as a function 
of redshift for GRB hosts collected by \cite{2009ApJ...691..182S} 
(large blue dots) and GOODS South galaxies by \cite{2006A&A...449..951G} 
(small black dots). 
Circled large blue dots show dark GRB hosts defined by optical-to-X-ray spectral 
index of $\beta_{\rm OX} < 0.5$ 
\citep{2004ApJ...617L..21J,2009RAA.....9.1103Z,2009ApJ...699.1087V}. 
GRB 080325 host is shown with a red square. Stellar masses in this figure 
are based on \cite{2003ApJ...593..258B} IMF, which is adopted for the 
Savaglio et al. sample. 
}
\label{figure9}
\end{center}
\end{figure}

Figure \ref{figure9} shows 
stellar masses of GRB hosts by \cite{2009ApJ...691..182S} (large blue dots), 
GOODS South galaxies by \cite{2006A&A...449..951G} (small black dots), 
and GRB 080325 host (a red dot) against redshift.
Figure \ref{figure9} clearly demonstrates that GRB 080325 host is one of 
the most massive GRB hosts 
($M_{*,\rm BG}$ = 7.0 $\times$ 10$^{10} M_{\odot}$, described below) among 
dark and non-dark GRB hosts previously identified. 
We note that the methods of the SED fitting analysis are, 
in some degree, different between GRB hosts by 
\cite{2009ApJ...691..182S} and other samples including 
GOODS South galaxies and GRB 080325 hosts. 
\cite{2009ApJ...691..182S} assumed the IMF by 
\cite{2003ApJ...593..258B} and two stellar-population components, 
while our SED-fitting analysis for GRB 080325 
host and GOODS South galaxies is based on Salpeter IMF and a single 
stellar-population component as described in \S 
\ref{MASSIVE RED HOST GALAXY AT}. Thus, to examine 
the effect by using different methods, 
stellar masses of GRB hosts (except for hosts for which only two-band data are available)
by \cite{2009ApJ...691..182S} were compared 
with those derived by our method. 
After the systematic difference between two IMFs is taken into account, 
i.e., a total stellar mass derived by  
Salpeter IMF is 1.8 times larger than that derived by using 
\cite{2003ApJ...593..258B} IMF, 
we found that both stellar masses agree with each other within 1 $\sigma$ uncertainty of 
$\delta \log M_{*} = 0.4$. 
In Figure \ref{figure9}, $M_{*,\rm Salp}$ of the GRB 080325 host and GOODS-South field galaxies 
are converted to $M_{*,\rm BG}$ which is stellar mass derived by 
\cite{2003ApJ...593..258B} IMF.

\begin{figure}
\begin{center}
\includegraphics[scale=0.7,angle=-90]{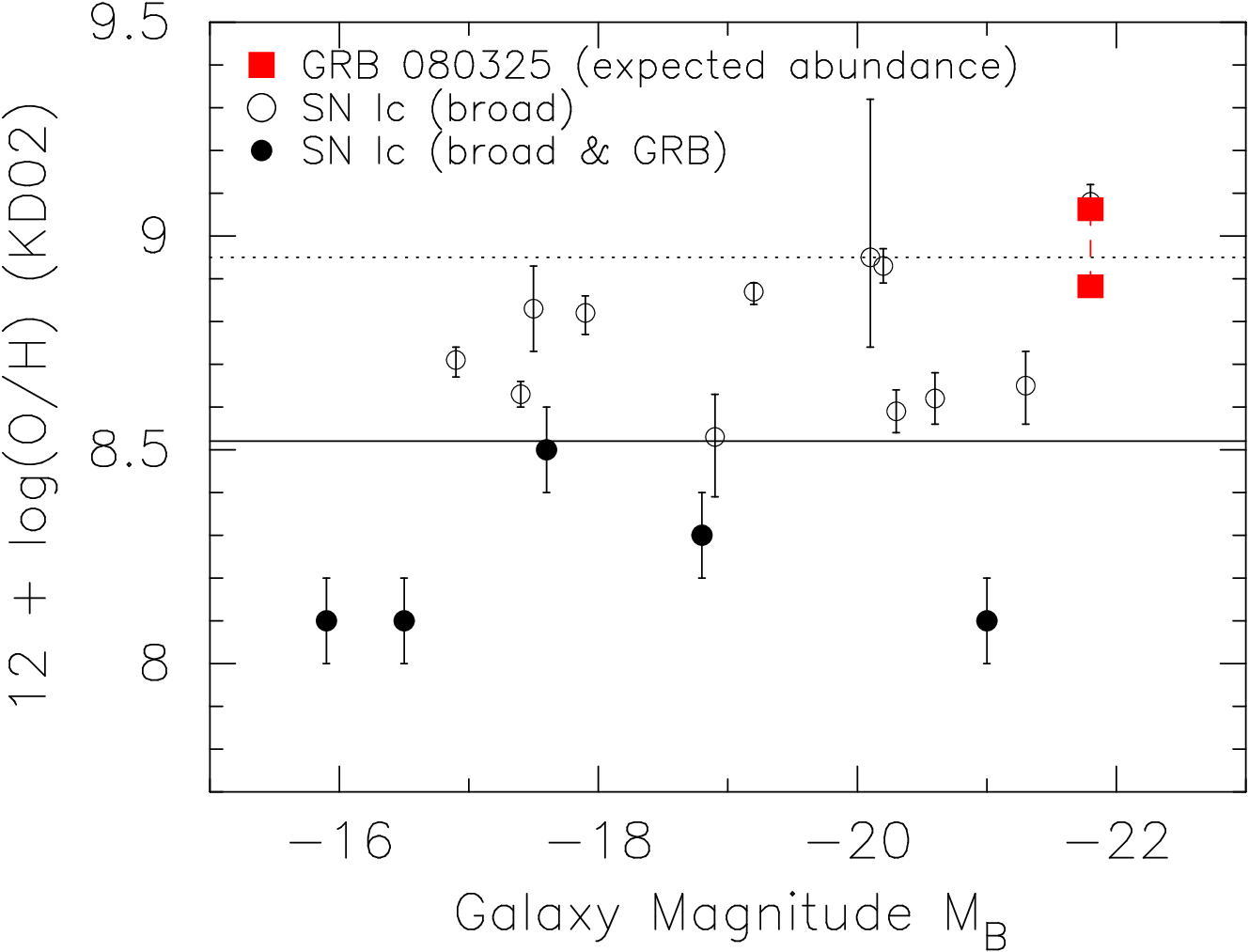}
\caption{
Metallicity of the GRB 080325 host, expected from mass-metallicity 
relations by \cite{2006ApJ...644..813E} and \cite{2009ApJ...691..140H} 
(red squares connected by a dashed line) is compared with the critical 
metallicity proposed by \citep{2008AJ....135.1136M}. 
Filled circles represent type Ic SNe 
with GRBs, and open circles are type Ic SNe without GRBs.
The critical and solar abundances are shown by solid and dotted 
lines, respectively. 
Expected metallicity of the GRB 080325 host is calibrated to 
\cite{2002ApJS..142...35K} method using conversion coefficients 
listed in \cite{2008ApJ...681.1183K}.
}
\label{figure10}
\end{center}
\end{figure}

Since the stellar mass of the GRB 080325 host is large, the metallicity may exceed the 
critical metallicity suggested by \cite{2008AJ....135.1136M} 
below which GRBs ($z < 0.14$) occur. 
If we assume the mass-metallicity relations obtained for 
star-forming galaxies at $z$ $\sim$ 2 \citep{2006ApJ...644..813E, 2009ApJ...691..140H}, 
the expected oxygen abundances (hereafter referred as metallicity) 
based on \cite{2002ApJS..142...35K} method are 12 + log (O/H) = 8.88 or 9.06. 
This expected metallicity of GRB 080325 host significantly 
exceeds the critical metallicity as shown in Figure \ref{figure10}. 
This suggests a high metallicity environment around GRB 080325. 

Similarly, high metallicity environments of GRBs were recently reported for 
GRB 020819 and GRB 051022 by spectroscopic observations 
\citep{2010ApJ...712L..26L,2009AIPC.1133..269G}. 
These GRBs commonly have massive hosts and are classified as dark. 
Thus, these three cases including GRB 080325 suggest the dusty high metallicity environment of GRBs. 
As one explanation, \cite{2010ApJ...712L..26L}
suggest the possibility that the enhanced mass loss rates 
associated with higher metallicities could potentially contribute 
to large amount of circumburst extinction, i.e., to produce a dark GRB. 

On the other hand, numerical 
calculations for a single star scenario of the GRB phenomenon suggest the 
low-metallicity 
environment around GRB progenitors is required \citep{2006ARA&A..44..507W,2006A&A...460..199Y}. 
Thus the possible high metallicity environment around GRB 080325 seems to favor 
binary-star merger scenario \citep{1995PhR...256..173N,1999ApJ...526..152F,2000ApJ...534..660I}. 
In this scenario, the merging stars take in their orbital 
angular momentum during the merging process. Therefore, the resulting collapsar 
can maintain a sufficient angular momentum required for a GRB explosion 
even if the high-metallicity environment. 
However, we note that 
the metallicity at the GRB position may be smaller than the critical metallicity, because of 
the metallicity gradient in the host (the position of GRB 080325 is at the edge of the host). 


\section{SUMMARY}
\label{SUMMARY}
We successfully detected an afterglow of GRB 080325 in $K_{s}$ band 
at 8.7 hours after the burst and its host galaxy in $K_{s}$ and 
$J$ band with Subaru/MOIRCS although no optical counterpart was reported. 
GRB 080325 is classified as a \lq\lq Dark GRB\rq\rq based on 
optical-to-X-ray spectral index of $\beta_{\rm OX} < 0.2-0.5$. 
The flux ratio of rest-wavelength optical to X-ray bands of the afterglow 
shows a very large dust extinction along the line of sight to the afterglow 
($A_{V}$ = $2.7 - 10$ mag). 
To reveal host properties, we obtained optical images in $B$, $R_{c}$, $i$', and $z$' 
bands with Subaru/Suprime-Cam at about one year after the burst, and 
clearly detected the host galaxy in these bands. 
The SED fitting analysis for the host galaxy was performed assuming 
the Calzetti et al. extinction law and various 
star formation histories including instantaneous, exponentially declining 
($e^{-\frac{t}{\tau}}$ with $\tau$ = 1 Myr, 10 Myr, 100 Myr,1 Gyr, and 10 Gyr), 
and constant star formation rate. The best-fit stellar population synthesis model 
($\tau$ = 1 Gyr) indicates that the host is at $z_{photo} =$ 1.9. The dust extinction 
for the entire host ($A_{V} = 0.8$ mag) is larger than those of typical GRB hosts. 
Although GRB 080325 is located at the outskirt of the host galaxy, the dust extinction 
along the line of sight to GRB 080325 is larger than that for the entire host. 
Therefore, a major reason for the optical faintness of GRB 080325 
relative to X-ray brightness is likely attributed to the local dusty environment around 
GRB 080325 rather than extinction by dust distributed over the entire host. 
We found that the host is very luminous (comparable to $L^{*}$ at $z \sim 2.0$) 
and very massive ($M_{*} = 7.0 \times 10^{10} M_{\odot}$) in contrast to the 
faint and less massive properties of GRB hosts at lower redshifts.
Considering the mass-metallicity relation for star-forming galaxies at $z \sim 2$ 
and the stellar mass of the GRB 080325 host, the high metallicity environment around GRB 080325 
is suggested. This possible high metallicity environment favors the binary merger 
scenario of the GRB phenomenon rather than the single star explosion, 
although the future spectroscopic observation at the GRB position is essential. 

\acknowledgments
We are grateful to the Subaru Telescope staffs for conducting ToO and Service 
observations. We would also like to thank Fumiaki Nakata 
and Sakurako Okamoto for useful comments on data reduction 
and analysis. 
We also thank Marcin Sawicki for providing us the \lq\lq {\it SEDfit} \rq\rq. 
We would also like to thank the anonymous referee for his/her comments 
which improved the paper. 
This work is supported by the Grant-in-Aid for Scientific Research Priority Areas 
(19047003) and the Grant-in-Aid for the Global COE Program "The Next Generation 
of Physics, Spun from Universality and Emergence" 
from the Ministry of Education, Culture, Sports, Science and Technology (MEXT) of Japan.

\clearpage

\end{document}